\documentclass[twocolumn,10pt,aps,prb,amsmath,amssymb,floatfix,citeautoscript,longbibliography]{revtex4-2}

\usepackage{graphicx}
\usepackage[caption=false]{subfig}
\usepackage{amsmath}
\usepackage{amssymb}
\usepackage{bm}
\usepackage{xcolor}
\usepackage{upgreek}
\usepackage[colorlinks, citecolor={blue!80!black}, urlcolor={blue!80!black}, linkcolor={red!50!black}]{hyperref}
\usepackage{bookmark}
\usepackage[version=4]{mhchem}
\usepackage{microtype}
\usepackage{colordvi}

\newcommand{\comment}[1]{}

\renewcommand{\comment}{\paragraph}

\begin{document}
\title{Spin relaxation in a single-electron bilayer graphene
quantum dot}

\author{Lin Wang}
\affiliation{Department of Physics, University of Konstanz, D-78457 Konstanz, Germany}
\author{Guido Burkard}
\affiliation{Department of Physics, University of Konstanz, D-78457 Konstanz, Germany}

\begin{abstract}
We study the spin relaxation in a single-electron bilayer graphene quantum dot due to the spin-orbit coupling. 
The spin relaxation is assisted by the emission of acoustic phonons via the bond-length change and deformation potential mechanisms and $1/f$ charge noise. In the perpendicular magnetic-field dependence of the spin relaxation rate $T_1^{-1}$, we predict 
a monotonic increase of $T_1^{-1}$ at higher fields where the electron-phonon coupling via the deformation potential plays a 
dominant role in spin relaxation. We show a less pronounced dip in $T_1^{-1}$ at lower magnetic fields due to the competition between the electron-phonon coupling due to bond-length change and $1/f$ charge noise. Finally, detailed comparisons of the magnetic-field dependence of the spin relaxation with the existing experiments by 
Banszerus \emph{et al.} [Nat. Commun. {\bf 13}, 3637 (2022)] and G\"achter \emph{et al.} [PRX Quantum {\bf 3}, 020343 (2022)] are reported.
\end{abstract}

\maketitle

\section{Introduction}
Bilayer graphene (BLG) allows for the formation of quantum dots (QDs), since the presence of an out-of-plane electric field opens a band gap~\cite{McCann2006,Min75, Castro99, Zhang2009,Konschuh85,McCann_2013} because the two constituent layers of graphene experience different electrostatic potentials. This enables BLG a possible candidate to form quantum dots (QDs) via electrostatically-induced quantum confinement~\cite{Trauzette2007,Banszerus2020, Banszerus2021,Banszerus13, Gachter3, Garreis2024,denisov2025spin,Banszerus2024}. The out-of-plane electric field also breaks the spatial inversion symmetry, and therefore the two inequivalent valleys in gapped BLG, i.e., the ${\bm K}$ and ${\bm K^{\prime}}$ points of the hexagonal Brillouin zone, have opposite Berry curvatures and associated orbital magnetic moments~\cite{Knothe98,Eich2018,Banszerus2020}. In the presence of an out-of-plane magnetic field, a valley splitting arises due to the valley Zeeman effect, which has already been 
demonstrated by single-carrier measurements in BLG QDs~\cite{Banszerus2021}. This valley Zeeman effect presents a promising route to control the valley degree of freedom and possibly establish valley-based qubits in BLG QDs~\cite{Rycerz2007,Rohling2012,Rohling2014,Schaibley2016}. Very recently, valley relaxation times, which directly limit the lifetime of the encoded information, have been reported to be remarkably long in both single-particle 
BLG QD~\cite{Banszerus2024} and singlet-triplet BLG double QDs~\cite{Garreis2024}.

Compared with valley, spin-based qubits in BLG QDs~\cite{Banszerus13, Gachter3} have received more attention because of the outstanding spin properties in BLG including 
low hyperfine interaction and weak spin-orbit coupling (SOC)~\cite{Kane95, Huertas74, Min74, Yao75, Boettger75, Fischer80, Gmitra80, Huertas103, Abdelouahed82, Konschuh82}. Recently, Banszerus \emph{et al.} measured spin relaxation 
times as long as $200\ \mu$s at a magnetic field of $1.9$~T in a single-electron BLG QDs~\cite{Banszerus13}. In addition, G\"achter \emph{et al.} reported the spin relaxation times up to $50\ $ms at a magnetic field of $1.7$~T~\cite{Gachter3}. To explain these two experiments, theoretical work on spin relaxation in BLG QDs is required.

\begin{figure}
\includegraphics[width=1.0\columnwidth]{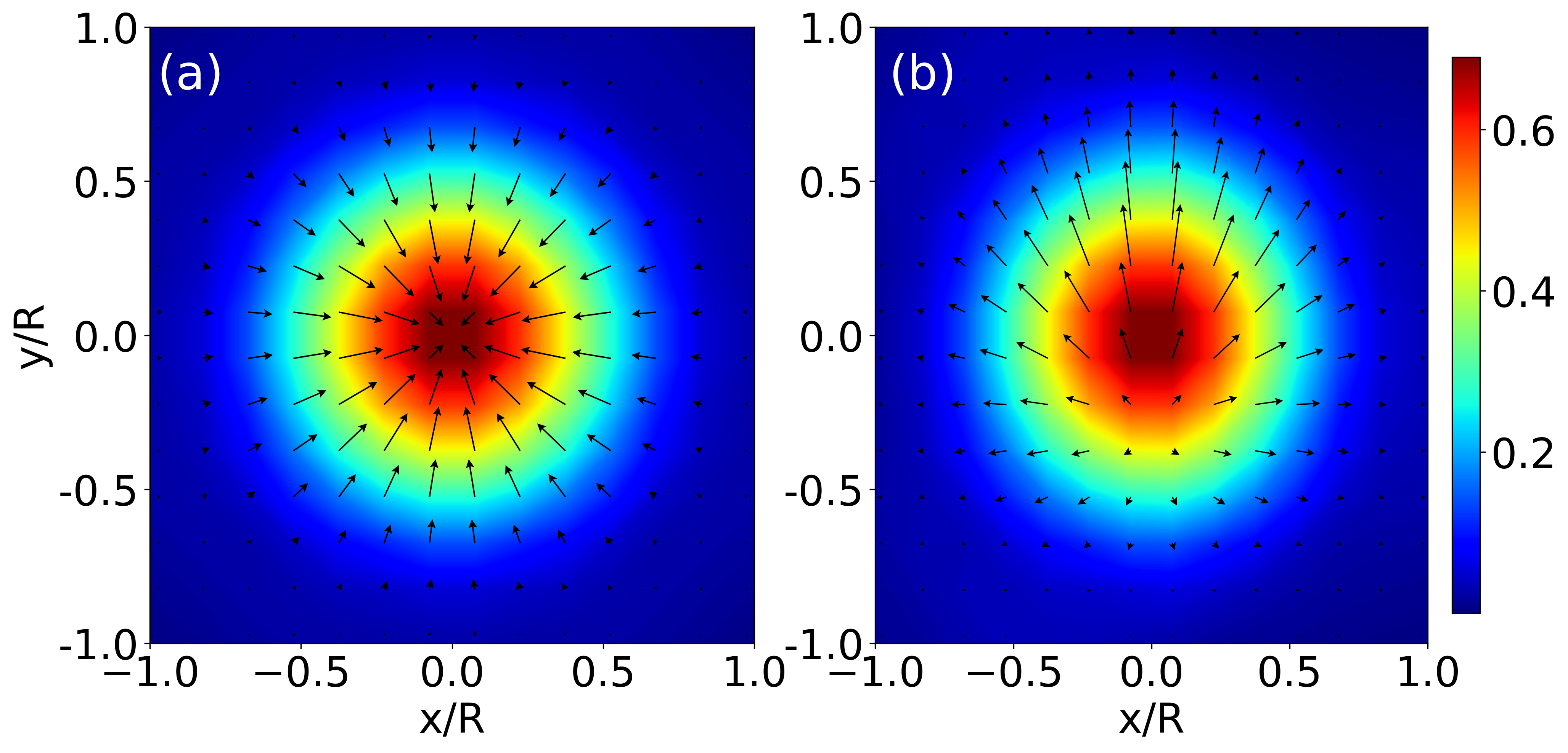}
\caption{Electron density (color scale, units of $1/R^{2}$) and in-plane spin fields (arrows) of the groundstate wavefunction of the BLG QD at $B_{\perp}=0.5\ $T and $E_z=0.5\ $V/nm with (a) $\gamma_3=0$ and (b) $\gamma_3\ne 0$. In the calculation, we have used $U_0=42.5\ $meV, $V=25\ $meV,  $R=25\ $nm.} 
\label{fig:spin_field}
\end{figure}

In this work, we study the spin relaxation time of a single-electron BLG QD as a function of the out-of-plane magnetic fields $B_\perp$. We first employ an exact diagonalization method to 
obtain the eigenvalues and the eigenstates and then calculate the spin relaxation time 
using Fermi's golden rule. The spin relaxation is induced by the spin-mixing due to SOC together with (i) electron-phonon coupling via the deformation potential and the bond-length change and (ii) $1/f$ charge noise. The spin relaxation rate is dominated by the electron-phonon coupling ($1/f$ charge noise) at higher (lower) fields. In the crossover between the low and high fields, a dip is predicted in the magnetic-field dependence arising from the competition between the electron-phonon coupling and the $1/f$ charge noise. Moreover, we show detailed comparisons with the recent experiments on spin relaxation times at higher magnetic fields~\cite{Banszerus13, Gachter3}. Good agreement is obtained between our numerical results and the experimental data.

This paper is organized as follows. In Section~\ref{sec:model} we present our model and the methods used to calculate the spin relaxation time. In Sec.~\ref{sec:mechanisms} we include the relevant spin relaxation mechanisms, and in Sec.~\ref{sec:results} we present the resulting spin relaxation rates obtained from our numerical calculation and then show the detailed comparison with the recent experiments. We summarize in Sec.~\ref{sec:conclusions}.

\section{Model and Method}
\label{sec:model}
With a homogeneous out-of-plane magnetic field $B_{\perp}$ 
and an electrostatic confinement potential 
$U({\bf r})$ with ${\bf r}=(x,y)$, the total single-particle Hamiltonian of Bernal (AB) stacked BLG is given by 
\begin{eqnarray}
H_{\rm QD}=H^{\tau}({\bf k})+U({\bf r})+H_{\rm SO}+H_{\rm Z}.\label{total_hamil}  
\end{eqnarray}
The first term $H^{\tau}({\bf k})$ represents a low-energy effective $4\times 4$ Hamiltonian near the ${\bm K}$ and ${\bm K^{\prime}}$ valleys \cite{Konschuh85,Wang87},
\begin{eqnarray}
H^{\tau}({\bf k})=\left(\begin{array}{cccc}
V & \gamma_0 p & \gamma_4 p^* & \gamma_1 \\
\gamma_0 p^* & V & \gamma_3 p & \gamma_4 p^* \\
\gamma_4 p & \gamma_3 p^* & -V & \gamma_0 p \\
\gamma_1 & \gamma_4 p & \gamma_0 p^* & -V
\end{array}\right),\label{eq5}
\end{eqnarray}
in the on-site orbital Bloch basis $\Psi_{\rm A_1}({\bf k})$, $\Psi_{\rm B_1}({\bf k})$,
$\Psi_{\rm A_2}({\bf k})$ and
$\Psi_{\rm B_2}({\bf k})$, where A and B refer to the sublattices, the subscripts $1,2$ denote two layers, and ${{\bf k}}=(k_x,k_y)$ stands for the two-dimensional
wave vector counted from the ${\bm K}$ ($\tau=1$) or ${\bm K^{\prime}}$ ($\tau=-1$) point.
Here,  the potential difference between two layers $2V=eE_zd_{\rm eff}$ is tunable by an out-of-plane electric field $E_z$ with $d_{\rm eff}$ being the effective electrostatic bilayer distance \cite{Konschuh85}, $\gamma_{0,1,3,4}$ represent the hopping parameters and the momentum dependence $p({\bf k})=-\sqrt{3}a(\tau {k}_x-i{k}_y-ixB_{\perp}e/2-\tau yB_{\perp}e/2)/2$ includes the orbital effect due to the out-of-plane magnetic 
field, with $a$ representing the lattice constant. The QD confinement potential is described by the second term in Eq.~\eqref{total_hamil}.  To make calculations feasible, we choose a circularly symmetric potential with finite step height. Specifically, when $r\ge R$, $U({\bf r})=U_0$; otherwise, $U({\bf r})=0$ with $U_0$ and $R$ being the potential depth and QD radius, respectively. The term 
$H_{\rm SO}=H_{\rm I}+H_{\rm BR}+H_{\rm inter}$ accounts for the SOC in BLG \cite{Konschuh85} where 
\begin{eqnarray}
H_{\rm I}&=&\Delta_{\rm I}\tau\sigma_z s_z,\\
H_{\rm BR}&=&\frac{1}{2}(\bar{\lambda}_0\mu_z+2\lambda_{\rm BR}\mu_0)(\tau\sigma_x s_y-\sigma_y s_x),\\
H_{\rm inter}&=&-\frac{1}{2}(\bar{\lambda}_4\sigma_z+\delta\lambda_4\sigma_0)(\tau\mu_x s_y+\mu_y s_x)\nonumber\\
&&\hspace{-0.5cm}\hbox{}+\frac{\lambda_3}{4}[\mu_x(\tau\sigma_x s_y+\sigma_y s_x)+\mu_y(\tau\sigma_y s_y-\sigma_x s_x)].
\end{eqnarray}
Here, $H_{\rm I}$ represents the intrinsic SOC; $H_{\rm BR}$ describes the Bychkov-Rashba-like SOC; $H_{\rm inter}$ stands for the terms due to interlayer coupling. $\Delta_{\rm I}$, $\bar{\lambda}_0$, $\lambda_{\rm BR}$, $\bar{\lambda}_4$, $\delta\lambda_4$, and $\lambda_3$ are the strengths of SOCs, and $\sigma_{x,y,z}$, $s_{x,y,z}$ and $\mu_{x,y,z}$ denote the Pauli matrices for the sublattice, spin and layer degree of freedom, respectively. The last term $H_{\rm Z}=g_s\mu_B s_zB_{\perp}/2$ describes the spin Zeeman coupling with $g_s$ the spin $g$-factor and $\mu_{B}$ the Bohr magneton. Note that the intervalley coupling is not included here since its 
contribution to spin mixing is negligible. All the parameters are listed in Table~\ref{tab:kp_mu}.
\begin{table}[b]
\caption{Parameters used in the calculation: $\gamma_{0,1,3,4}$, $d_{\rm eff}$, $g_s$, $a$, $\bar{\lambda}_0$, $\lambda_{\rm BR}$, $\bar{\lambda}_4$, 
$\delta\lambda_4$, $\lambda_3$ and $\Delta_{\rm I}$ are introduced in Eq.~(\ref{total_hamil}), $v_{\rm TA,LA}$, $g_{1,2}$ and $\rho$ are used in Eq.~(\ref{ele_phonon}). Note that $E_z^*$ is the magnitude of $E_z$ in units of V/nm.}
\begin{tabular}{c c  c | c c}
\hline\hline
$\gamma_0$ & $2.6\ $eV &\quad\quad & $\gamma_1$ & $0.339\ $eV \\
$\gamma_3$ & $0.28\ $eV & & $\gamma_4$ & $-0.14\ $eV \\
$g_s$ & $2$ & & $a$ & $2.46\ \textup{\AA}$\\
$v_{\rm TA}$ & $1.22\times 10^4\ $m/s & & $v_{\rm LA}$ & $1.95\times 10^4\ $m/s\\
$\bar{\lambda}_0$ & $5\ \mu$eV & & $\rho$ & $1.52\times 10^{-7}\ $g/cm$^{-2}$\\
$\bar{\lambda}_4$ & $-12\ \mu$eV & & $\lambda_{\rm BR}$ & $5E_z^*\ \mu$eV\\
$\delta\lambda_{4}$ & $-3E_z^*\ \mu$eV & & $\lambda_{3}$ & $1.5E_z^*\ \mu$eV\\
$\Delta_{\rm I}$ & $43\ \mu$eV & & $g_1$ & $50.0\ $eV\\
$g_2$ & $2.8\ $eV & & $d_{\rm eff}$ & $0.1\ $nm\\ 
\hline\hline
\end{tabular}
\label{tab:kp_mu}
\end{table}

In the absence of the intervalley coupling, two valleys are independent. For each valley, we first solve the Schr\"{o}dinger equation 
of the Hamiltonian $H_0=H^{\tau}({\bf k})+U({\bf r})$ numerically by applying a discretization in real space by introducing a two-dimensional square
grid. To avoid the Fermi doubling problem that arises from lattice discretization, a Wilson mass
term $wk^2$ is taken into account in $H_0$ with $w$ denoting the Wilson mass~\cite{Messias96}. Then, we include all other terms in Eq.~(\ref{total_hamil}) under the eigenbasis of $H_0$ with both spin up and down. Finally, 
single-particle eigenvalues and eigenstates can be obtained by exactly diagonalizing the total Hamiltonian.

\section{Spin relaxation mechanisms}
\label{sec:mechanisms}
The spin relaxation rate between the initial state $|i\rangle$ and the final state $|f\rangle$ can be calculated using Fermi's golden rule by taking into account (I) the electron-phonon 
coupling via the bond-length change and the deformation potential mechanisms \cite{Struck2010,Droth2011,Droth2013,wangvalley}, and (II) $1/f$ charge noise \cite{Hosseinkhani2021,Hosseinkhani2022,wangvalley}.

\subsection{\footnotesize Relaxation due to electron-phonon coupling}
In the present work, two different electron-phonon interaction mechanisms are considered, i.e., the deformation potential arising from an area change in the unit cell induced by phonons and the bond-length change due to a modified hopping matrix element. Regarding the phonon modes, we only take into account the acoustic 
phonons near the $\Gamma$ point since we are interested in the low-energy regime. Among all acoustic phonon modes, out-of-plane ones are irrelevant by assuming that the BLG sheet is grown on a substrate. As a result, only in-plane longitudinal-acoustic (LA) and transversal-acoustic (TA) 
modes are considered. The intravalley electron-phonon coupling in the sublattice basis can be written as \cite{Ando2005},
\begin{align}
H_{\rm EPC}^{\lambda q}=\frac{q}{\sqrt{A\rho\Omega_{{\bf q},\lambda}}}\left(\begin{array}{cc}
g_1a_1     &g_2a^*_2  \\
g_2a_2     &g_1a_1 
\end{array}\right)(e^{i{\bf q}\cdot{\bf r}}b_{\lambda {\bf q}}^{\dagger}-e^{-i{\bf q}\cdot{\bf r}}b_{\lambda {\bf q}}).
\label{ele_phonon}
\end{align}
Here, $b_{\lambda {\bf q}}^\dagger$ and $b_{\lambda {\bf q}}$ denote the creation and annihilation operators for branch $\lambda={\rm TA},{\rm LA}$ phonons with wavevector ${\bf q}$, $A$ is the area of the graphene sheet, $\rho$ represents the mass density of BLG, $\Omega_{{\bf q}, \lambda}=v_{\lambda}q$ describes the phonon energy with $v_{\lambda}$ the sound velocity, $g_1$ ($g_2$) stands for the coupling strength of the deformation 
potential (bond-length change), $a_1=i$ and $a_2=ie^{2i\phi_{\bf q}}$ for LA phonons, and $a_2=e^{2i\phi_{\bf q}}$ and 
$a_1=0$ for TA phonons.Note that we assume the electron-phonon coupling to be the same for both graphene layers by considering the weak interlayer coupling.

Using Fermi's golden rule, we can calculate the spin relaxation rate due to electron-phonon coupling from $|i\rangle$ with energy $\epsilon_i$ to $|f\rangle$ with energy $\epsilon_f$ as 
$T_1^{-1}=2\pi A\sum_{\lambda}\int\frac{d^2q}{(2\pi)^2}|\langle i|H_{\rm EPC}^{\lambda q}|f\rangle|^2\delta(\epsilon_f-\epsilon_i+\Omega_{{\bf q},\lambda})$.
Here, by assuming that the temperature is much lower than the spin splitting, we only take into account the phonon emission process.

\subsection{\footnotesize Relaxation due to \texorpdfstring{$1/f$}{} charge noise}
In the environment of the localized QD electron, fluctuating two-level systems often cause the 
electric (charge) noise with its typical $1/f$ power spectral density. $S_{\rm E}(\omega)=S_0/\omega^{\alpha}$ describes the electric charge noise spectra where $S_0$ represents the 
power spectral density at $1\ $Hz and the exponent $\alpha$ is device dependent and typically between $0.5$ and $2$~\cite{Kranz32}. The spin relaxation rate from $|i\rangle$ to $|f\rangle$ due to $1/f$ charge noise can be calculated by $T_1^{-1}={4\pi e^2}/{{\hbar}^2}S_{\rm E}(\epsilon_i-\epsilon_f)\sum_j|\langle i|r_j|f\rangle|^2$ 
with ${\bf r}=(x, y)$ \cite{Hosseinkhani2021,Hosseinkhani2022}.

\begin{figure}
\includegraphics[width=1.0\columnwidth]{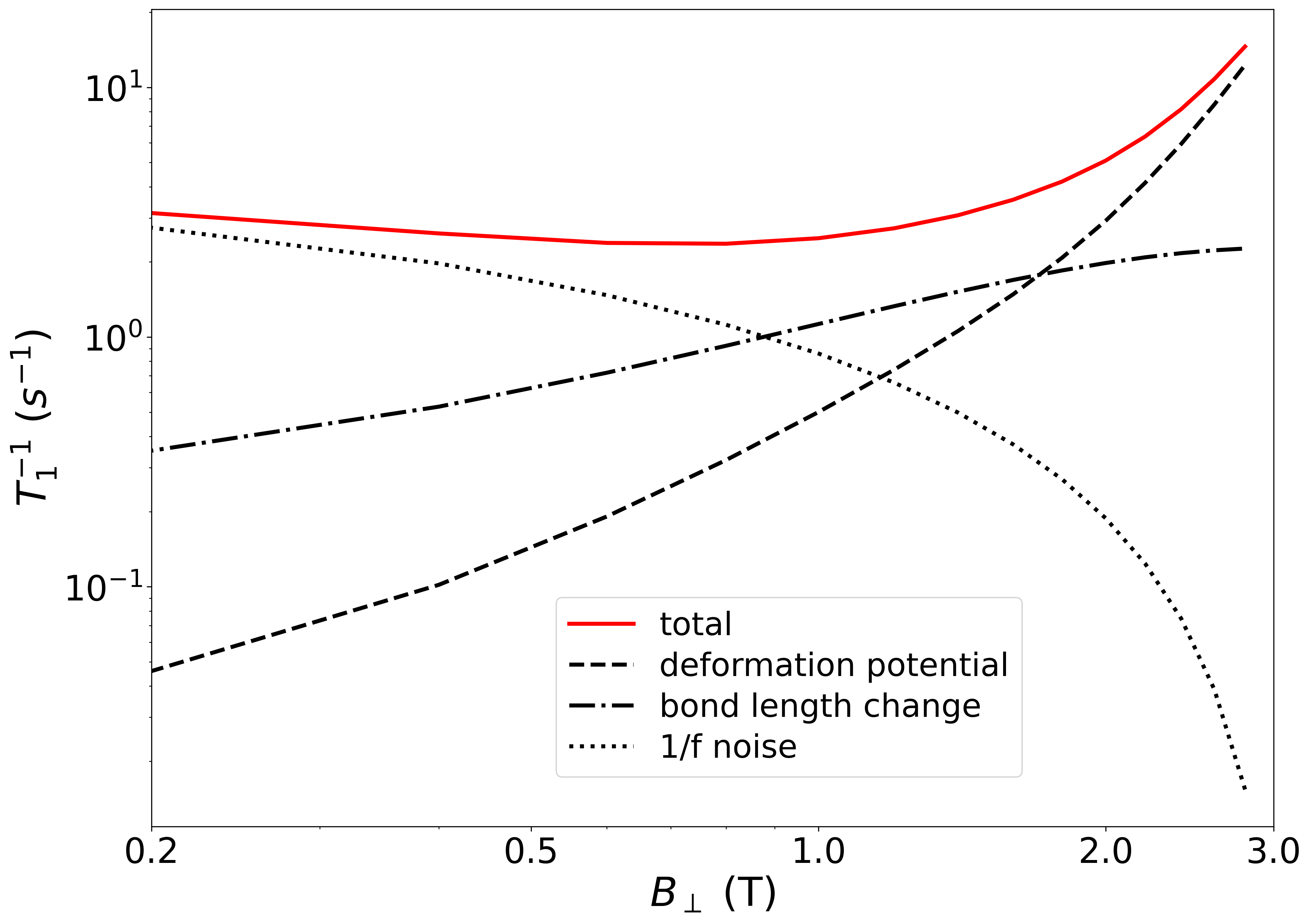}
\caption{Total spin relaxation rate $T_1^{-1}$ (red curve) and spin relaxation rate $T_1^{-1}$ due to deformation potential (dashed curve), bond-length change (dotted-dashed curve) or $1/f$ charge noise (dotted curves), as a function of perpendicular field $B_{\perp}$ on a double-logarithmic scale. In the calculation, we have used $U_0=42.5\ $meV, $V=25\ $meV,  $R=25\ $nm, $S_0=20\ \mu$eV$^2/$Hz, $\alpha=0.8$ and $E_z=0.5\ $V/nm.} 
\label{fig:T1_theory}
\end{figure}

\section{Numerical results}
\label{sec:results}
In Fig.~\ref{fig:spin_field}, we plot the electron density of the groundstate wavefunction of a BLG QD at $B_{\perp}=0.5\ $T as a function of the coordinates $x$ and $y$. We find that the wavefunction is well localized within the QD. In addition, we also show the in-plane spin texture~\cite{luo2019unique} by arrows for an electric field $E_z=0.5\ $V/nm with $\gamma_3=0$ and $\gamma_3\ne 0$. In the absence of $\gamma_3$, the spin texture is invariant under rotations~\cite{wangvalley}. However, with $\gamma_3$ taken into account, the spin texture becomes anisotropic since the rotational symmetry is broken. 

\begin{figure}[t]
\includegraphics[width=1.0\columnwidth]{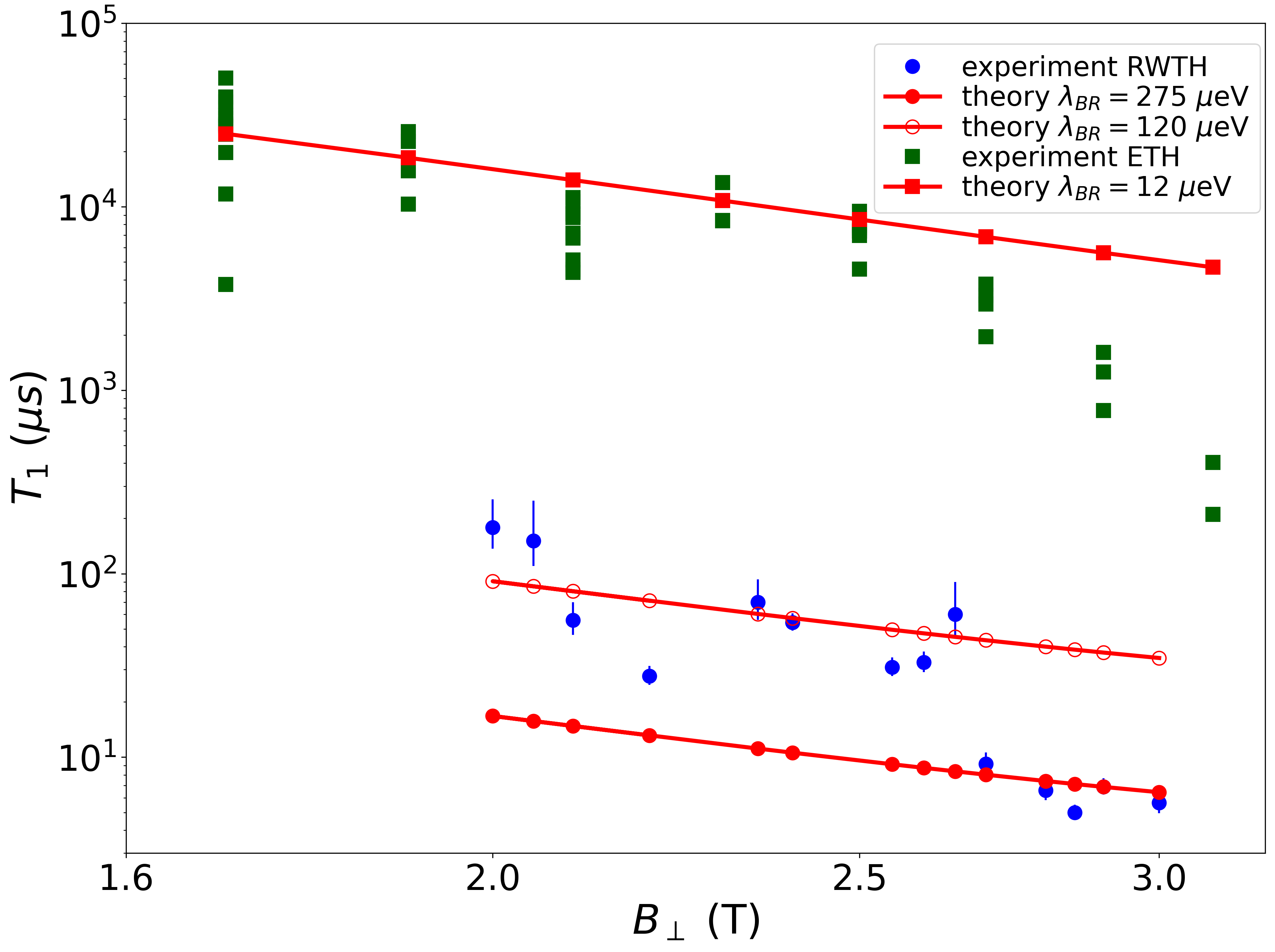}
\caption{Log-log plot of the dependence of the spin relaxation time $T_1$ on the perpendicular magnetic field $B_{\perp}$. The dark green squares and the blue dots represent the experimental data from the ETH Z{\"u}rich group~\cite{Gachter3} and from the RWTH Aachen group~\cite{Banszerus13}, respectively. 
The red curve with squares (dots) denotes the theoretical results obtained by fitting to the experimental data from the ETH Z{\"u}rich 
(RWTH Aachen) group using least square method with fitting parameter $\lambda_{\rm BR}=12\ \mu$eV ($\lambda_{\rm BR}=275\ \mu$eV). 
Open circles are obtained from a least-squares fit without error bars and with $\lambda_{\rm BR}=120\ \mu$eV.
Other parameters are $U_0=42.5\ $meV, $V=25\ $meV,  $R=25\ $nm. Note that all SOC parameters 
except $\lambda_{\rm BR}$ are calculated with $E_z=0.5\ $V/nm.} 
\label{fig:T1_exp}
\end{figure}

\subsection{Spin relaxation: theoretical calculation}
Without the intervalley coupling, two valleys $K$ and $K^{\prime}$ are independent. Then, we can calculate the spin relaxation rate within each valley. Here, throughout the paper, we focus on the spin relaxation rate between the first excited state and the ground state within $K^{\prime}$ valley.

In Fig.~\ref{fig:T1_theory}, we plot the total spin relaxation rate as a function of the perpendicular magnetic field $B_\perp$. The total spin relaxation rate first decreases and then increases with the increase of the magnetic field. To understand this behavior, we show the spin relaxation rate due to deformation potential, 
bond length change, and $1/f$ charge noise separately in Fig.~\ref{fig:T1_theory}. At higher fields, the spin relaxation rate is dominated by the electron-phonon coupling, in particular, the deformation potential. The spin relaxation rate due to the electron-phonon coupling increases with increasing magnetic field because the spin splitting between initial and final states becomes larger. This leads to a monotonic increase of spin relaxation rate at higher fields. However, at lower fields, the spin relaxation rate due to $1/f$ charge noise plays a more important role, resulting in the decrease of the spin relaxation rate with increasing magnetic field. In the crossover between the low and high fields, a dip is predicted in the magnetic-field dependence due to the competition between $1/f$ noise and electron-phonon coupling.

\subsection{Comparison with experiments}
Two reports on experimental measurements of the single-particle spin relaxation in BLG QDs are available, one by Banszerus~\emph{et al.}~\cite{Banszerus13} from RWTH Aachen, and another by G\"{a}chter~\emph{et al.}~\cite{Gachter3} from ETH Z{\"u}rich, plotted as blue dots and dark green squares, respectively, in Fig.~\ref{fig:T1_exp}. For both experiments, the spin relaxation times present a behavior of overall decay with increasing magnetic field. However, the spin relaxation time from ETH is about two or three orders of magnitude larger than that from RWTH. To explain these two experiments, the contribution of deformation potential and bond length change is taken into account. At the same time, $1/f$ charge noise is not considered since its contribution at higher fields is negligible as shown in Fig.~\ref{fig:T1_theory}. Since there are too many parameters in our model, it is very expensive to treat them all as fitting parameters. In the present work, we only treat one SOC parameter $\lambda_{\rm BR}$ as fitting parameter for simplicity. We perform a least-squares fit to the experimental data. The result of our numerical fit to the experiment from ETH Z{\"u}rich group is shown as the red curve with squares with $\lambda_{\rm BR}=12\ \mu$eV. Our numerical fit with (without) error bars to the experiment from RWTH Aachen group is plotted as the red curve with (open) circles with $\lambda_{\rm BR}=275\ \mu$eV ($\lambda_{\rm BR}=120\ \mu$eV). Our results agree with the experimental data both qualitatively and quantitatively. Note that at very high fields, an observable difference is shown between the experimental data and our numerical results. This requires further research on spin relaxation 
both experimentally and theoretically.

\section{Conclusions}
\label{sec:conclusions}
We have investigated the electronic spin relaxation in a BLG QD in the presence of a perpendicular magnetic field $B_\perp$. The spin relaxation arises from spin mixing due to spin-orbit coupling together 
with electron-phonon coupling via bond-length change and deformation potential mechanisms and $1/f$ charge noise. 
At higher fields, the spin relaxation rate is dominated by the electron-phonon coupling, leading to a monotonic increase with increasing magnetic field. However, at lower fields, $1/f$ charge noise becomes more important, which results in the decrease of the spin relaxation rate. In the crossover between the low and high fields, we predict a dip in the magnetic-field dependence arising from the competition between the contributions of electron-phonon coupling and $1/f$ noise. In addition, we show our comparison with the recent experiments of Banszerus~\emph{et al.}~\cite{Banszerus13} and G\"{a}chter~\emph{et al.}~\cite{Gachter3}. Our results agree well with the experimental data both qualitatively and quantitatively.

\acknowledgments
We acknowledge Katrin Hecker, Luca Banszerus, and Wei Wister Huang for fruitful discussions.

\bibliography{spin_T1}
\end{document}